\documentclass[dvipsnames]{article}
\usepackage[margin=1in]{geometry}
\usepackage{graphicx,amsmath,upgreek,bm}
\usepackage{colortbl}
\usepackage{wrapfig}
\usepackage[rgb, table]{xcolor}
\usepackage{booktabs}
\usepackage{cite}
\usepackage{makecell}
\usepackage{pifont}
\usepackage{todonotes}
\usepackage[toc,page]{appendix}
\usepackage{multirow}
\usepackage{algorithm}
\usepackage{rotating}
\usepackage{hyperref}
\hypersetup{
   colorlinks=true,
   linkcolor=blue,
   filecolor=magenta,
   urlcolor=blue,
   citecolor=blue,
	}
\usepackage[
  separate-uncertainty = true,
  multi-part-units = repeat
]{siunitx}
\usepackage{caption} 
\usepackage{subcaption}
\usepackage{amsmath}
\usepackage{gensymb}
\usepackage{mathtools}
\usepackage{amssymb}  
\usepackage{appendix}    
\usepackage{booktabs} 
\usepackage[para,online,flushleft]{threeparttable}
\usepackage{textcomp}
\usepackage[noend]{algpseudocode}
\usepackage{stackengine}

\usepackage{arydshln}
\makeatletter
  \renewcommand*\env@matrix[1][*\c@MaxMatrixCols c]{%
    \hskip -\arraycolsep
    \let\@ifnextchar\new@ifnextchar
  \array{#1}}
\makeatother
\usepackage{graphicx}
\usepackage{textcomp}
\usepackage{xcolor}
\usepackage{float}
\usepackage{stackengine}
\usepackage{multicol}
\usepackage{makecell}
\usepackage{colortbl}
\usepackage{stfloats}

\makeatletter
\def\BState{\State\hskip-\ALG@thistlm}
\makeatother
\definecolor{light-gray}{gray}{0.9}
\newcommand{\Z}{\mathbb{Z}}

\title{Chatter Diagnosis in Milling Using Supervised Learning and Topological Features Vector}

\usepackage{authblk}

\author[1]{Melih C. Yesilli}
\author[2]{Sarah Tymochko}
\author[1]{Firas A. Khasawneh}
\author[2,3]{Elizabeth Munch}
\affil[1]{Department of Mechanical Engineering, Michigan State University}
\affil[2]{Computational Mathematics, Science and Engineering, Michigan State University}
\affil[3]{Department of Mathematics, Michigan State University}

\date{}

\begin{document}
\maketitle

\begin{abstract}
Chatter detection has become a prominent subject of interest due to its effect on cutting tool life, surface finish and spindle of machine tool.
Most of the existing methods in chatter detection literature are based on signal processing and signal decomposition.
In this study, we use topological features of data simulating cutting tool vibrations, combined with four supervised machine learning algorithms to diagnose chatter in the milling process.
Persistence diagrams, a method of representing topological features, are not easily used in the context of machine learning, so they must be transformed into a form that is more amenable.
Specifically, we will focus on two different methods for featurizing persistence diagrams, Carlsson coordinates and template functions.
In this paper, we provide classification results for simulated data from various cutting configurations, including upmilling and downmilling, in addition to the same data with some added noise.
Our results show that Carlsson Coordinates and Template Functions yield accuracies as high as $96\%$ and $95\%$, respectively.
We also provide evidence that these topological methods are noise robust descriptors for chatter detection.
\end{abstract}


\section{Introduction}
\label{sec:intro}
Productivity in discrete manufacturing processes, such as turning and milling, is often constrained due to the occurrence of harmful, large amplitude oscillations called chatter.
Although there is active research on chatter predictive models \cite{Quintana2011} as well as methods for chatter diagnosis and mitigation \cite{Munoa2016,Yilmaz2002,Al-Regib2003,Dijk2010}, in-process chatter detection remains a challenging task.
Some of the factors that complicate chatter identification include: the complexity of the cutting process which involves several interacting systems, the presence of nonlinearities and noise, and the shift in the process parameters during cutting.
This necessitated the search for tools for chatter identification from sensors instrumented to the cutting center.
The output of these sensors is often a time series, which is an equi-spaced record of a physical quantity such as acceleration versus time.

The standard approach for chatter recognition from cutting signals has mostly focused on extracting features by decomposing the time series and combining them with supervised learning algorithms---most commonly Support Vector Machine (SVM).
The two most widely used decompositions are Wavelet Packet Transform (WPT) \cite{Yao2010,Chen2017,Saravanamurugan2017,Qian2015,Xie2016,Cao2013,Choi2003,Yoon2005,Yesilli2019a} and Ensemble Empirical Mode Decomposition (EEMD) \cite{Yesilli2019a,Ji2018,Chen2018,Liu2011,Cao2015}.
Both WPT and EEMD  require manually preprocessing the signal to identify the most informative parts of the signal which carry chatter signatures, which is characterized by the part of the decomposition whose spectrum contain the chatter frequency.
Once the informative decompositions are obtained, they are used to compute several time and frequency domain features for chatter classification.
Many times the resulting features are too many and they overfit the model; therefore, the traditional tools are often equipped with a feature ranking process to prune the features' vector \cite{Chen2017,Chen2018,Yesilli2019a}.

However, one of the main challenges with decomposition-based methods is that they require a new training set for every cutting configuration where the latter can simply be a variation in the mounting of the workpiece of the tool on the same cutting center.
Since changes in the configuration can change the system's eigenfrequencies, the corresponding chatter frequencies will also shift yielding any previously extracted informative decompositions inaccurate for the new configuration.
This requires a dedicated skilled user to examine the signal for each configuration, tag chatter and chatter free cases, and extract the informative decompositions.
In other words, classifiers trained using WPT or EEMD features can be difficult to generalize to different cutting conditions even for the same machine \cite{Yesilli2019a}.
Recently, topological features were also explored for chatter detection in turning with the advantage of bypassing the manual feature extraction phase inherent to WPT and EEMD \cite{Yesilli2019}.
The mathematical model for turning is often a delay differential equation with constant coefficients.
However, despite the promising success that the TDA tools showed in turning, their viability for other subtractive manufacturing processes such as milling has not been tested yet.
The mathematical treatment of milling models is significantly more difficult than turning because they are often described using (linear) delay differential equations with time periodic coefficients.
Therefore, the interplay between the delay and the periodicity leads to more complicated behavior.
Further, the transition from chatter-free to chatter in milling is not so simple, and it can occur for certain perturbations due to the existence of unsafe zones characterized by unstable quasi-periodic oscillation which limit the basin of attraction of the chatter-free stable periodic motion \cite{Dombovari2019}.
Chatter can even manifest as chaotic motion in some cases, further complicating its identification \cite{Szalai2004}.

In this paper, we combine features obtained with TDA with supervised machine learning algorithms to diagnose chatter in milling.
The data used in this study is obtained from simulating the oscillations of a single degree of freedom milling tool with four straight cutting teeth \cite{Tlusty2000}.
We consider both upmilling and downmilling processes.
The classification robustness of the described approach to noise is demonstrated by adding noise with several signal-to-noise ratio to the data.
The time series are then embedded into point clouds using Takens embedding theorem \cite{Takens1981} where the embedding parameters are obtained using a permutation entropy based method \cite{Myers2019}.
Features are then extracted from the 1D and 2D persistence diagrams of the point cloud using two different featurizations of the persistence diagram: Carlsson coordinates \cite{Adcock2016} and template functions \cite{Perea2019}.
In addition to studying using TDA for chatter versus chatter-free classification, we further investigate the ability of the approach to identify the bifurcation associated with chatter, specifically: chatter-Hopf, chatter-period doubling, and chatter-free.
We employ the four most common traditional classification algorithms: Support Vector Machine (SVM), Logistic Regression (LR), Random Forest (RF) and Gradient Boosting (GB).
The tagging of the simulated time series is obtained from the stability diagram which was computed using the spectral element method \cite{Khasawneh2011}.
The stability diagram marks the chatter and chatter-free regions in the process parameter space, i.e., the spindle speed and the depth of cut \cite{Bobrenkov2010}.

Our results show that for the investigated process featurization via template functions outperforms its Carlsson coordinates counterpart in most of the cases.
A comparison of the success of 1D versus 2D features in identifying chatter shows that the combination both 1D and 2D persistence provides the best average accuracy for both feature extraction methods while only using 2D features deteriorates the classification accuracy.

This paper is organized as follows.
Section \ref{sec:milling_simulation} explains the milling process model and the data labeling.
Section \ref{sec:TDA} provides background information on TDA.
Section \ref{sec:Featurization} explains how the features are extracted from the persistence diagrams, while
Section \ref{sec:Results} compares and discusses the classification results.

\section{Modeling}
\label{sec:milling_simulation}
\begin{figure}[hbt!]
\centering
\includegraphics[width=0.50\textwidth,height=0.7\textheight,keepaspectratio]{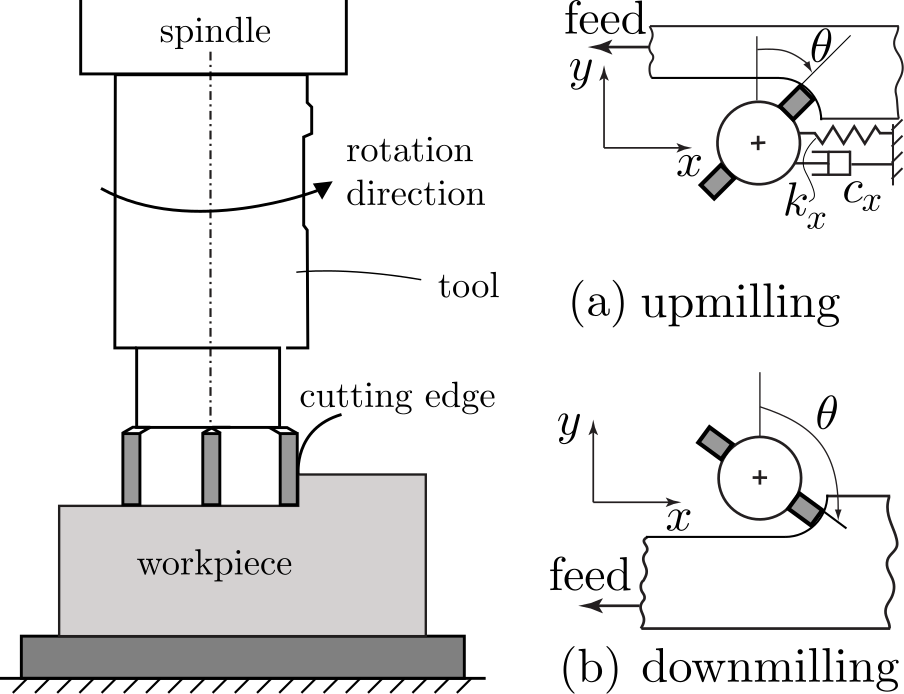}
\caption{Milling process illustrations. a) Upmilling b) Downmilling}
\label{fig:up_down_milling}
\end{figure}

We consider a milling operation with straight edge cutters as shown in Fig.~\ref{fig:up_down_milling}. 
A single degree of freedom model in the $x$ direction for the tool oscillations is used as shown in Fig.~\ref{fig:up_down_milling}a, and both upmilling and downmilling processes are considered in the analysis.  
The equation of motion that describes the tool oscillations is 
\begin {equation}
\label{eq:eom_end_mill}
\ddot{x}+ 2\zeta \omega_{n}\dot{x}+\omega_{n}^{2} x = \frac{1}{m} F(t),
\end{equation}
where $m$, $w_{n}$, $\zeta$ and $F(t)$ represent the modal mass, natural frequency, damping ratio and the cutting force in the $x$ direction, respectively. 
$\tau$ is the time delay given by $\tau = 2\pi/N\Omega$ where $\omega$ is the spindle's rotational speed in rad/s, while $N$ is the number of cutting edges or teeth. 
The expression for the cutting force is given by \cite{Altintas2009,Bobrenkov2010}
\begin{equation}
\label{eq:modified_cutting_force}
F= \sum\limits_{n=1}^z \big[-bK_{t} g_n(t)( \cos{\theta_{n}(t)}+ \tan{\gamma} \sin{\theta_{n}(t)}) \sin{\theta_{n}(t)}\big]
\big[(f+x(t)-x(t-\tau))\big],
\end{equation}
where $\theta_{n}$ is the angle between the vertical line and the leading tooth of the cutting tool as shown in Fig.~\ref{fig:up_down_milling}. 
The constant $K_{t}$ is the linearized cutting coefficient in the tangential direction and $\tan{\gamma}=K_{n}/K_{t}$ where $K_n(t)$ is the cutting coefficient in the normal direction. 
The screening function $g_{n}(t)$ is either $0$ or $1$ depending on whether the $n$th tooth is engaged in the cut or not, respectively, and $f$ represents the feed per tooth of the cutting tool. 
The expression for angular position of the $n$th tooth $\theta_{n}(t)$ is given by \cite{Bobrenkov2010}
\begin{equation}
\theta_{n}(t) = (2\pi\Omega/60)t+2\pi(n-1)/z,
\end{equation}
where $z$ is the total number of cutting teeth while $\Omega$ is the rotational speed given in revolutions per minute (rpm).

One of the important cutting parameters is the radial immersion ratio (RI) which is defined as the ratio of the radial depth of cut to the diameter of the cutting tool. 
Smaller radial immersions indicate shallower cuts and thus more intermittent contact between the tool and the workpiece, while higher radial immersions indicate deeper cuts with a more continuous contact.  
In our simulations for both downmilling and upmilling we set $RI=0.25$. 

Inserting Eq.~\eqref{eq:modified_cutting_force} into Eq.~\eqref{eq:eom_end_mill} results in
\begin{equation}
\label{eq:modified_eom_milling}
\ddot{x}(t)+2\zeta \omega_{n}(t)\dot{x}(t)+w_{n}^{2}x(t)=
-\frac{bh(t)}{m}[x(t)-x(t-\tau)]-\frac{bf_{0}(t)}{m},
\end{equation}
where $b$ is the nominal depth of cut and  $h(t)$ is the $\tau$-periodic function 
\begin{equation}
h(t)=\sum\limits_{n=1}^z K_{t} g_n(t) \big[  \cos{\theta_{n}(t)}+\\ \tan{\gamma} \sin{\theta_{n}(t)} \big]  \sin{\theta_{n}(t)},
\end{equation}
and $f_0(t)=h(t)\, f$. 
The term $f_0(t)$ does not affect the stability analysis, so we drop it in the subsequent equations; however we keep it in the simulation. 

\begin{figure}[hbt!]
\centering
\includegraphics[width=0.40\textwidth,height=0.7\textheight,keepaspectratio]{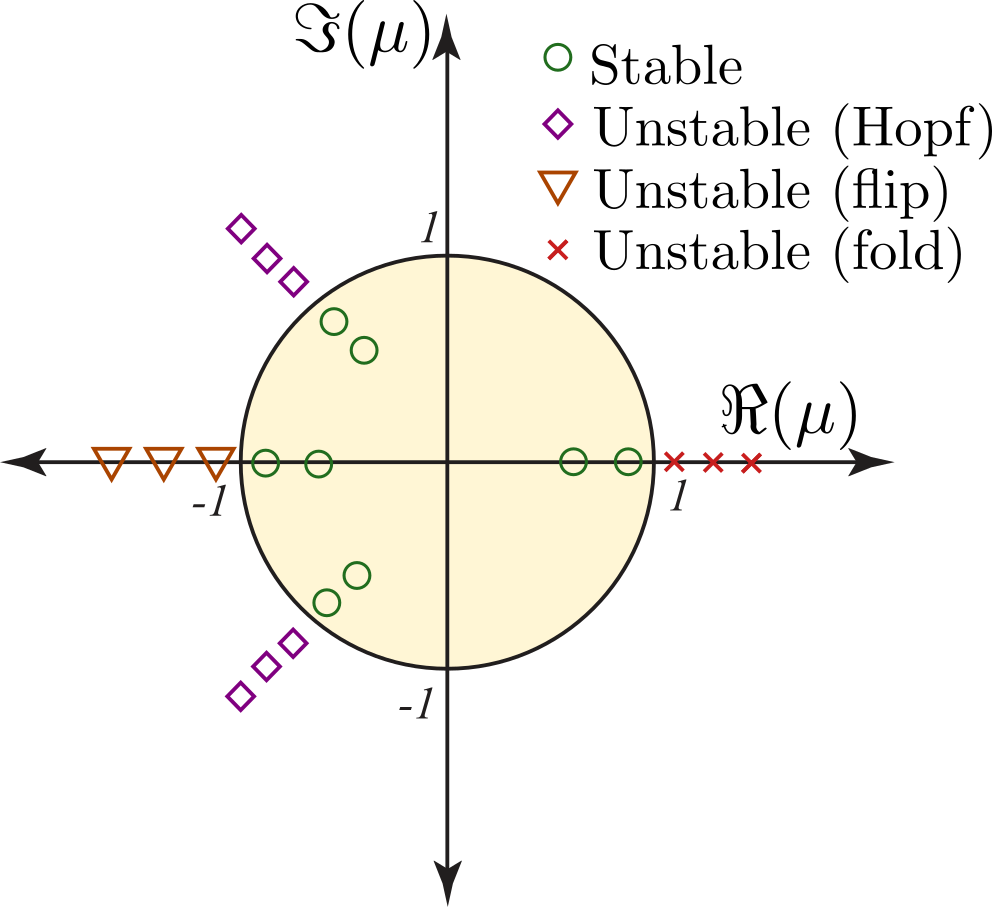}
\caption{Stability criteria used in this study based on the eigenvalues of the monodromy matrix $\textbf{U}$.}
\label{fig:stability_circle}
\end{figure}

After dropping $f_0(t)$, the equations of motion can be written in state space form according to
\begin{equation}
\frac{d\boldsymbol{\xi}(t)}{dt} = \textbf{A}(t)\boldsymbol{\xi}(t)+\textbf{B}(t)\boldsymbol{\xi}(t-\tau),
\end{equation}
where $\textbf{A}$ and $\textbf{B}$ are $T$-periodic with $T=\tau$. 
Then, using the spectral element method \cite{Khasawneh2011}, the state space is discretized and we obtain the dynamic map
\begin{equation}
\boldsymbol{\xi}_{n+1} = U\boldsymbol{\xi}_{n},
\end{equation}
where $\textbf{U}$ is the finite dimensional monodromy operator.  
The eignenvalues of $U$ approximate the eigenvalues of the infinite dimensional monodromy operator of the equation of motion. 
If the modulus of the largest eigenvalue is smaller than $1$, then the corresponding spindle speed and depth of cut pair leads to a chatter-free process; otherwise, chatter occurs. 
Therefore, the stability of the milling model and the bifurcation associated with the loss of stability (chatter) can be obtained by examining these eigenvalues, see Fig.~\ref{fig:stability_circle}. 

In this study, we generated $10000$ time series corresponding to a $100\times100$ grid in the plane of the spindle speeds and depths of cut. 
Each time series is tagged using the largest eigenvalue of the monodromy matrix corresponding to the same grid point. 

\section{Topological Data Analysis}
\label{sec:TDA}
Topological Data Analysis (TDA) extracts information from the data by quantifying its shape and structure.
One of the main tools of TDA is persistent homology.
Specifically, in this paper we study the time series by embedding them using delay reconstruction and then applying $1$-D persistent homology to obtain an information-rich summary of the shape of the resulting ambient space.
Features are then extracted from the persistence diagrams and used for machine learning.
This section briefly describes the main concepts of $1$-D persistent homology, but we defer a more thorough treatment to references in the literature such as \cite{Ghrist2008,Carlsson2009,Edelsbrunner2009,Oudot2015,Munch2017}.

\subsection{Simplicial complexes}
\label{sec:SimplicialComplexes}
An abstract $k$-simplex $\sigma$ is a subset of $k+1$ vertices, $\sigma \subset V$, whose dimension is given by $\dim(\sigma)=k$.
Some examples include the $0$-simplex which is a point, the $1$-simplex which is an edge, the $2$-simplex which is a triangle, and so on.
A simplicial complex $K$ is a set of simplices $\sigma  \in K$ that satisfies specific inclusion relations.
Specifically, if $\sigma \in K$, then all the lower dimensional component simplices  $\sigma' \subset \sigma$, called the faces of $\sigma$, are also in $K$; i.e.~$\sigma' \in K$.
For example, if a $2$-simplex (triangle) is in a simplicial complex $K$, then so are the corresponding $1$-simplices (edges of the triangle) as well as all the $0$-simplices (vertices of the triangle).


\subsection{Homology}
Assume that a simplicial complex $K$ is fixed, then the corresponding homology groups, denoted $H_n(K)$, can be utilized to quantify the holes of the structure in different dimensions.
For instance, the rank of the $0$ dimensional homology group $H_0(K)$ is the number of connected components (dimension $0$).
The rank of the $1$ dimensional homology group $H_1(K)$ is the number of loops (dimension $1$), while the rank of $H_2(K)$ is the number of voids (dimension $2$), and so on.

We will first explicitly construct the homology groups.
Given a simplicial complex $K$, the $n$-simplices of $K$ can be used as a generating set of the $\Z_2$-vector space $C_n(K)$ called the $n$th chain group.
In this representation, an element of $C_n(K)$ can be written as a finite formal linear combination $\sum_{\sigma \in K^{(n)}}{\alpha_{\sigma} \sigma}$, where $\alpha_\sigma \in \Z_2$.
Such an element is called an $n$-chain,  and addition of elements is accomplished by adding their coefficients.

We are now ready to define boundary operators.
Given a simplicial complex $K$, the boundary map $\partial_n: C_n(K) \to C_{n-1}(K)$ is defined on the generators by
\begin{equation*}
\partial_n([v_0, \ldots, v_n]) = \sum\limits_{i=0}^n{[v_0, \ldots, \hat{v}_i, \ldots, v_n]},
\end{equation*}
where $\hat{v}_i$ denotes the absence of element $v_i$ from the set, and is then extended linearly to be defined on all $n$-chains.
This linear transformation maps any $n$-simplex to the sum of its codimension $1$ faces.

By combining boundary operators, we obtain the chain complex
\begin{equation*}
\ldots \xrightarrow{\partial_{n+1}} \Delta_n(K) \xrightarrow{\partial_n} \ldots \xrightarrow{\partial_1} \Delta_1(K) \xrightarrow{\partial_0} 0,
\end{equation*}
with the fundamental property that the composition of any two subsequent boundary operators is zero, i.e., $\partial_{n} \circ \partial_{n+1}=0$.
An $n$-chain $\alpha \in \Delta_n(K)$ is a cycle if $\partial_n(\alpha) = 0$; it is a boundary if there is an $n+1$-chain $\beta$ such that $\partial_{n+1}(\beta) = \alpha$.
Define the kernel of the boundary map $\partial_n$ using $Z_n(K)=\{c \in \Delta_n(K): \partial_n c = 0\}$, and the image of $\partial_{n+1}$ $B_n(K)=\{c \in \Delta_n(K): c=\delta_{n+1} c', c' \in \Delta_{n+1}(K)\}$.
Consequently, we have $B_k(K) \subseteq Z_k(K)$.
Therefore, we define the $n$th homology group of $K$ as the quotient group $H_n(K)=Z_n(K)/B_n(K)$.
In this paper, we only need $1$- and $2$-dimensional persistent homology.
In the case of $1$-dimensional homology, there is one homology class in $H_1(K)$ for each hole in the complex.
For $2$-dimensional homology, there is only one homology class in $H_2(K)$ for each 2-dimensional void in the complex.
%

\subsection{Persistent homology}
\label{ssec:persistent_homology}
Homology is extremely useful for studying the structure of a simplicial complex.
However, it is limited to a static complex, but we are often interested in studying the structure of a changing simplicial complex.
For example, assume we have a point cloud $P \subset \mathbb{R}^m$, which in our case likely results from embedding a time series into $\mathbb{R}^m$, e.g., using delay reconstruction.
For each point $p \in P$, let $B(p,r)$ be the ball centered at $p$ and of radius $r$.
For each choice of the radius $r$ we can build a different simplicial complex with vertex set equal to the points.
The intersection of any two radius $r$ balls adds an edge between the associated vertices, the intersection of three balls adds a triangle, and higher dimensional analogs are added similarly.
For a specific choice of $r$, the resulting simplicial complex of the union of all the balls $\bigcup_{p\in P}B(p,r)$, called the \v{C}ech complex, gives a simplicial complex with the same topological properties as the union of balls, but it's construction is computationally prohibitive.

Luckily, the \v{C}ech complex is well-approximated by the Vietoris-Rips complex which is
\begin{equation*}
  \mathrm{VR}(r) = \{ \sigma \subseteq P \mid d(u,v) \leq r \, \forall u,v \in \sigma\}.
\end{equation*}
Notice that for any $r \leq s$, $\mathrm{VR}(r) \subseteq \mathrm{VR}(s)$.
If we let $\{r_1 < r_2 < \ldots < r_{\ell}\}$  be the set of the sorted distances between points, then the Rips filtration corresponding to the set of points $P$ is the ordered sequence of subcomplexes
\begin{equation*}
  \emptyset \subseteq K_1 \subseteq K_2 \subseteq \cdots \subseteq K_{\ell}
\end{equation*}
where $K_i = \mathrm{VR}(r_i)$ for the sake of notation.
This filtration enables the investigation of the structure of the points under multiple values for $r$ rather than a user-defined choice of proximity parameter.

The main idea behind persistent homology is to watch how the homology changes over the course of this given filtration.
Fix a homology dimension $n$, then any given filtration
\begin{equation*}
  K_1 \subseteq K_2 \subseteq \cdots \subseteq K_N
\end{equation*}
induces a sequence of linear maps on the homology
\begin{equation*}
  H_n(K_1) \to H_n(K_2) \to \cdots \to H_n(K_N).
\end{equation*}
We say that a class $[\alpha] \in H_n(K_i)$ is born at $i$ if it is not in the image of the map $H_n(K_{i-1}) \to H_n(K_i)$.
The same class dies at $j$ if $[\alpha] \neq 0$ in $H_n(K_{j-1})$ but $[\alpha] = 0$ in $H_n(K_{j})$.

\begin{figure}[h]
\centering
\includegraphics[width=0.65\textwidth,height=.8\textheight,keepaspectratio]{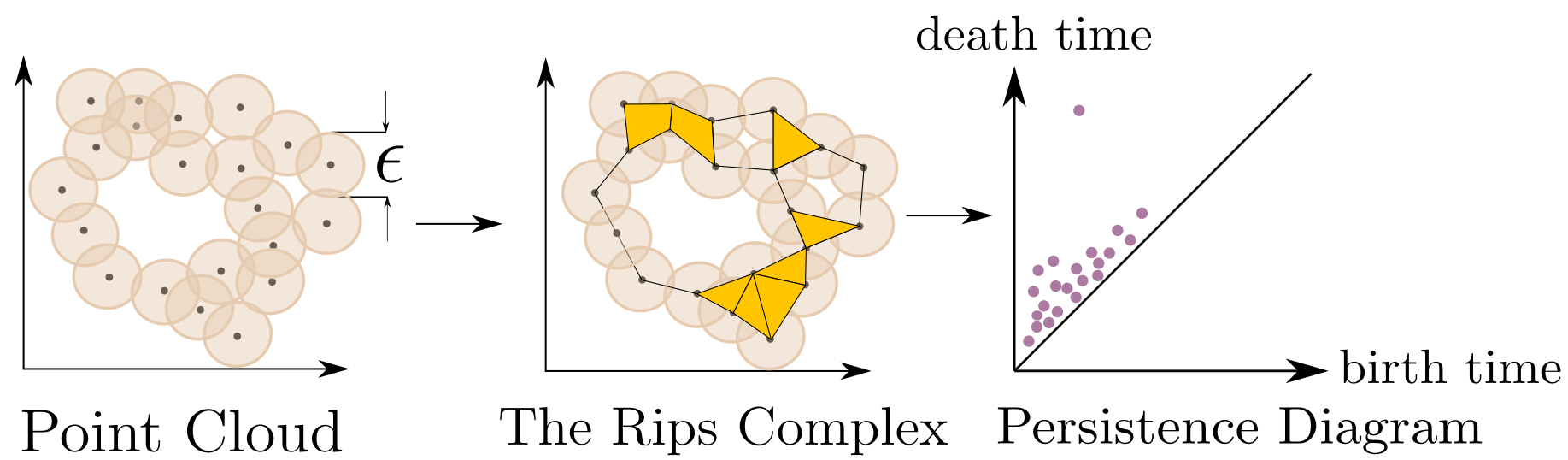}
\caption{The Rips complex.}
\label{fig:Rich_Cech_Complex}
\end{figure}

This information can be used to construct a persistence diagram $X$ as follows.
A class that is born at $i$ and dies at $j$ is represented by a point in $\mathbb{R}^2$ at $(i,j)$.
The collection of the points in the persistence diagram, therefore, gives a summary of the topological features that persist over the defined filtration.
We denote the number of the off-diagonal points in the persistence diagram by $|X|$.
See Fig.~\ref{fig:Rich_Cech_Complex} for an example of point cloud data, one step in the Rips filtration, and the resulting persistence diagram for $n=1$.

\section{Feature Extraction Methods}
\label{sec:Featurization}

We convert each embedded time series into a persistence diagram, and then convert the persistence diagram to a feature vector using one of the two methods described below.


\subsection{Carlsson Coordinates}
Carlsson coordinates are one method of featurizing persistence diagrams based on ideas from commutative algebra \cite{Adcock2016}.
The idea is to utilize polynomial functions which are applied to the points in the persistence diagram but are oblivious to the order of the points in the diagram.
 In this paper we use five functions following \cite{Adcock2016,Khasawneh2018}.
Given a persistence diagram, $D = \{(b_i,d_i)_{i=1}^n\}$, where $b_i$ and $d_i$ are corresponding birth and death values, and $d_{\rm max}$ represents maximum persistence, the Carlsson coordinates of $D$ are calculated using the functions
\begin{equation}
\begin{array}{rl}
\label{eq:CC_equations}
f_1(D) & = \sum{b_i(d_i-b_i)}, \\
f_2(D) & = \sum{(d_{\rm max} - d_i)(d_i - b_i)},\\
f_3(D) & = \sum{b_i^2 (d_i - b_i)^4},  \\
f_4(D) & = \sum{(d_{\rm max} - d_i)^2 (d_i-b_i)^4},\\
f_5(D) & = \max\{ (d_i - b_i) \}.
\end{array}
\end{equation}

In this study, we use these five features for 1D and 2D persistence diagrams separately, in addition to generating feature matrices by concatenating features for both dimensions.
While generating the features matrices, we use all possible combinations of features, a total of $\sum\limits_{i=1}^5{\tbinom{5}{i}}$.

\subsection{Template Functions}

Originally proposed in \cite{Perea2019}, template functions are another method of featurizing persistence diagrams.
We will briefly introduce this method and an example template function system known as interpolating polynomials.
For a more detailed background see \cite{Perea2019}.

For simplicity, in this section we will define everything in terms of persistence diagrams in birth-lifetime coordinates; that is, a point $(b,d)$ in birth-death coordinates is instead represented as $(b,d-b)$ where $d-b$ represents how long a feature lived.
Now rather than having all points in the persistence diagram on or above the diagonal, we have all points in the upper half plane, $\mathbb{W} := \mathbb{R}\times \mathbb{R}_{>0}$.

A \emph{template function} is any continuous function on $\mathbb{R}^2$ that has compact support contained in the upper half plane, $\mathbb{W}$ \footnote{If we were working in birth-death coordinates, the function would be required to have support above the diagonal.}.
Given a persistence diagram $D$, a template function can be turned into a function $f:\mathbb{W}\to \mathbb{R}$ by evaluating the function on each point in the diagram and then summing over all values.
That is,
\[
v_f(D) = \sum_{x\in D} f(x).
\]
A \emph{template system} is a collection of template functions, $\mathcal{T}$ such that the functions defined on persistence diagrams $\mathcal{F}_\mathcal{T} = \{v_f: f\in \mathcal{T}\}$ separate points in the diagram.
In other words, given two diagrams $D_1, D_2$, there is a function $f\in \mathcal{T}$ such that $v_f(D_1)\neq v_f(D_2)$.
While a true template system contains an infinite number of functions, in \cite[Thm.~29]{Perea2019} it is proven that any function on persistence diagrams can be approximated by some finite subset of a template system, so selecting this finite subset gives a vectorization of the persistence diagram $(v_{f_1}(D), \ldots, v_{f_k}(D) )$.
In this paper, we will use one example of a template system, interpolating polynomials, as introduced in \cite{Perea2019}.

Given a mesh $\mathcal{A} = \{a_i\}_{i=0}^m\subset \mathbb{R}$, the Lagrange polynomial corresponding to $a_j$, $\ell_j^\mathcal{A}(x)$, is defined as
\[
\ell_j^\mathcal{A} (x) = \prod_{i\neq j} \frac{x - a_i}{a_j - a_i}.
\]
This function satisfies
\[
\ell_j^{\mathcal{A}} (a_k) =
\begin{cases}
1 & j=k \\
0 & \text{otherwise}.
\end{cases}
\]
Fixing two meshes, $\mathcal{A}\subset \mathbb{R}$ and $\mathcal{B} \subset \mathbb{R}_{>0}$ and coordinates $i,j$, the template function is defined as
\[
f(x,y) = \beta(x,y) \cdot | \ell_{i}^\mathcal{A}(x) \ell_{j}^\mathcal{B}(y) |
\]
where $\beta$ is a bump function to force the resulting polynomial to have compact support within a designated area.
In practice, we choose the mesh $\mathcal{A}\times\mathcal{B}$ to ensure this region encloses all points in the diagram, then the region for compact support is implicit and so $\beta$ need not be specified.

\section{Results and Discussions}
\label{sec:Results}
\begin{figure}[h]
\centering
\includegraphics[width=0.95\textwidth,height=.95\textheight,keepaspectratio]{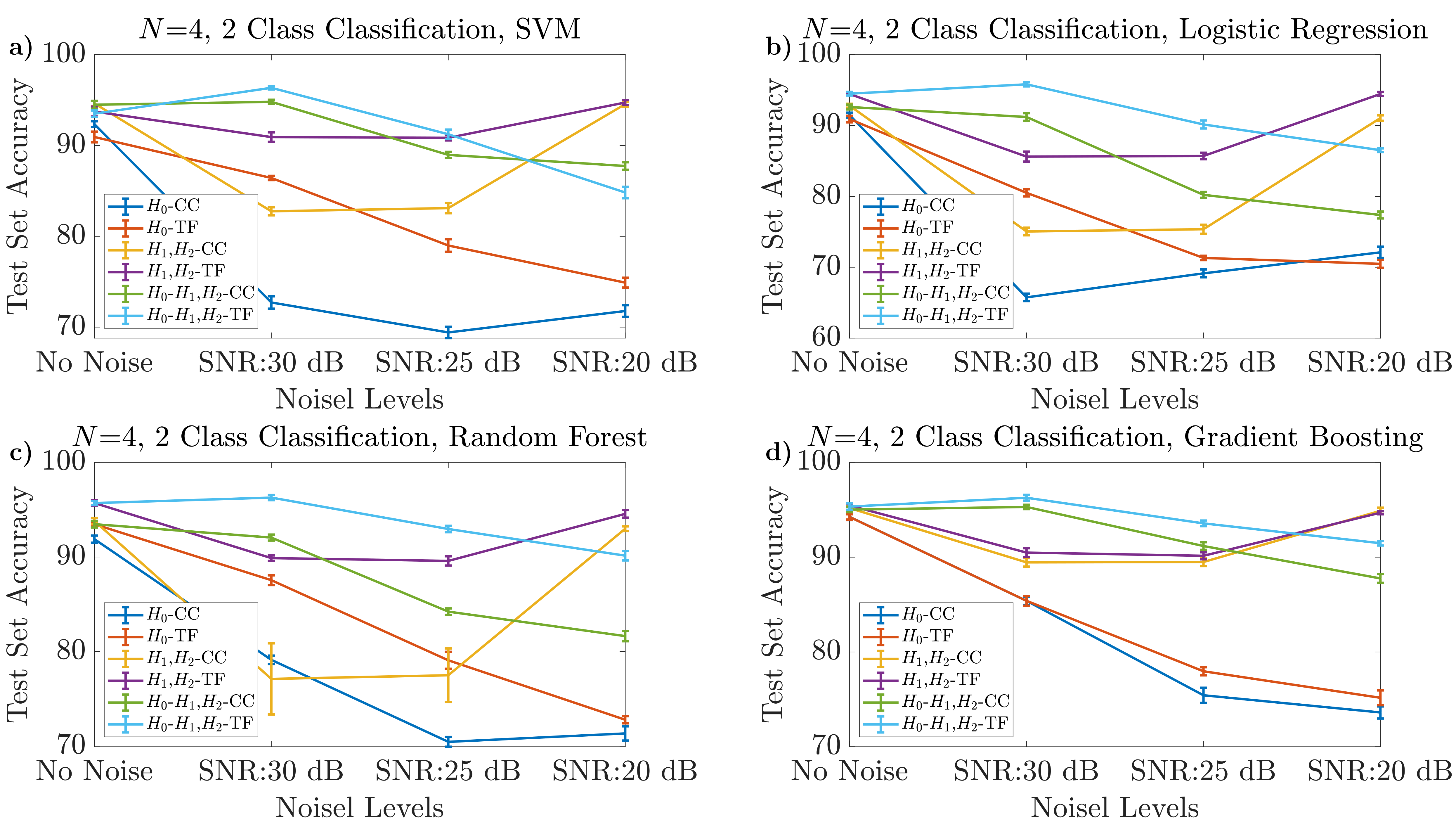}
\caption{Comparison of 0D, 1D and 2D persistence results obtained from downmilling data set with $N=4$. Results of a) Support Vector Machine, b) Logistic Regression, c) Random Forest Classifier, d)Gradient Boosting.}
\label{fig:H0_H1_H2_comp}
\end{figure}

In this section, we provide classification accuracies for each featurization method for noisy and non-noisy time series of upmilling and downmilling processes with $4$ teeth ($N=4$). 
Ranges of rotational speed and depth of cut parameters for the simulations are chosen with respect to the stability diagrams given for both processes in \cite{Bobrenkov2010}. 
The 1- and 2-dimensional persistence diagrams were used with the methods described in Section \ref{sec:Featurization}. 
Feature matrices were computed for 1D and 2D persistence diagrams individually, and the features were concatenated when using both dimensions. 0D persistence has been omitted in this study due to its poor performance on noisy data sets as evidenced by a reduction in the classification accuracy by $10\%$ in some cases as shown in Fig.\ref{fig:H0_H1_H2_comp}.
Data sets are randomly split, using $67\%$ for training and $33\%$ for testing. 
The split-train-test is performed $10$ times, and the mean accuracies with the corresponding standard deviations are reported in this section.

This data can be used for both a two class and three class classification problem.
The first is classifying either chatter-free or chatter, while the second further divides chatter into two types: Hopf-unstable and period2-unstable.
Classification for both two and three class problems is done using four different algorithms: support vector machines, logistic regression, random forests and gradient boosting.
Default parameters have been used for all classification algorithms except random forest classification ($n\_estimator=100$ and $max\_depth=2$).
These two types of chatter are based on Hopf and period doubling bifurcation behaviors as described in Fig.~\ref{fig:stability_circle}.

Two class classification results for downmilling and upmilling simulations with $N=4$ are provided in Tables~\ref{tab:Down_N4_CC_TF_2Class_Results} and \ref{tab:Up_N4_CC_TF_2Class_Results}, respectively.
For each data set, the highest accuracy is highlighted in blue. 
For instance, $95.5\%$ accuracy is obtained as the best classification accuracy for non-noisy data sets when gradient boosting classifiers are trained with combined 1D and 2D persistence features based on Template Functions method in Table~\ref{tab:Down_N4_CC_TF_2Class_Results}.
In most of the cases for downmilling and upmilling, it is seen that the highest accuracies are obtained when 1D and 2D persistence diagrams features are combined. 

Some of the time series embeddings especially in the chatter-free regime, do not have any $2$ dimensional topological features, thus giving an empty $H_2$ diagram.
If a specific cutting configuration has a lot of time series with empty $H_{2}$, feature matrices for these time series have a lot of zeros when either featurization method is used.
Because of the lack of $2$ dimensional information for many of the time series, classifications using only $H_2$ have lower accuracies than only using $H_1$ as is shown in Tables~\ref{tab:Down_N4_CC_TF_2Class_Results} and \ref{tab:Up_N4_CC_TF_2Class_Results}.

When comparing persistence diagram featurizations, the template function method has the best results for all data sets, with the exception of two: the noisy data set with SNR value of 20 dB for downmilling and the one without noise for upmilling. 
However, for those two data sets, template functions' results are very close to those provided by Carlsson coordinates. 
When comparing classification algorithms, SVM yields the highest accuracies for five of the eight data sets, while gradient boosting yields the highest accuracies for the remaining three data sets.
\begin{table*}[t]
\centering
\caption{2 Class classification results for noisy (SNR:20,25,30 dB) and non-noisy data sets which belong to downmilling process with $N=4$ (N: Teeth Number, CC: Carlsson Coordinates, TF: Template Functions, SVM: Support Vector Machine, LR: Logistic Regression, $H_{1}$: 1D persistence, $H_{2}$: 2D persistence).}
\label{tab:Down_N4_CC_TF_2Class_Results}
\resizebox{\textwidth}{!} {  
\begin{tabular}{|l|c|c|c|c|c|c|c|c|c|c|c|c|}
\hline
\multicolumn{1}{|c|}{Down Milling} & \multicolumn{6}{c|}{\shortstack{Without Noise}} & \multicolumn{6}{c|}{\shortstack{SNR: 20 dB}} \\
\hline
\makecell{$N=4$} & \multicolumn{3}{c|}{CC} & \multicolumn{3}{c|}{TF} & \multicolumn{3}{c|}{CC} & \multicolumn{3}{c|}{TF}  \\
\hline
\makecell{Classifier} & $H_{1}$ & $H_{2}$& $H_{1}$-$H_{2}$ & $H_{1}$ & $H_{2}$& $H_{1}$-$H_{2}$& $H_{1}$ & $H_{2}$& $H_{1}$-$H_{2}$& $H_{1}$ & $H_{2}$& $H_{1}$-$H_{2}$\\
\hline
SVM		&$\SI{94.3}{\percent}$   & $\SI{85.1}{\percent}$   & $\SI{94.6}{\percent}$   & $\SI{92.9}{\percent}$   & $\SI{94.4}{\percent}$   & $\SI{93.7}{\percent}$   & $\SI{94.2}{\percent}$   & $\SI{92.5}{\percent}$& $\SI{94.6}{\percent}$   & $\SI{94.3}{\percent}$   & $\SI{94.8}{\percent}$   & $\SI{94.7}{\percent}$  \\
LR	  &$\SI{92.4}{\percent}$   & $\SI{84.3}{\percent}$   & $\SI{92.8}{\percent}$   & $\SI{93.9}{\percent}$   & $\SI{91.6}{\percent}$   & $\SI{94.5}{\percent}$   & $\SI{78.5}{\percent}$   & $\SI{78.3}{\percent}$& $\SI{91.1}{\percent}$   & $\SI{93.8}{\percent}$   & $\SI{93.5}{\percent}$   & $\SI{94.5}{\percent}$  \\
RF    &$\SI{93.6}{\percent}$   & $\SI{90.9}{\percent}$   & $\SI{93.8}{\percent}$   & $\SI{95.0}{\percent}$   & $\SI{94.1}{\percent}$   & $\SI{95.7}{\percent}$   & $\SI{92.4}{\percent}$   & $\SI{92.3}{\percent}$& $\SI{93.0}{\percent}$   & $\SI{94.3}{\percent}$   & $\SI{94.4}{\percent}$   & $\SI{94.6}{\percent}$  \\
GB    &$\SI{95.0}{\percent}$   & $\SI{93.5}{\percent}$   & $\SI{95.2}{\percent}$   & $\SI{94.7}{\percent}$   & $\SI{94.2}{\percent}$   & \cellcolor[rgb]{0.13,0.67,0.8}$\SI{95.5}{\percent}$   & $\SI{94.2}{\percent}$   & $\SI{93.7}{\percent}$& \cellcolor[rgb]{0.13,0.67,0.8}$\SI{94.9}{\percent}$   & $\SI{94.7}{\percent}$   & $\SI{94.4}{\percent}$   & $\SI{94.7}{\percent}$  \\
\hline
\multicolumn{1}{|c|}{Down Milling} & \multicolumn{6}{c|}{\shortstack{SNR: 25 dB}} & \multicolumn{6}{c|}{\shortstack{SNR: 30 dB}} \\
\hline
\makecell{$N=4$} & \multicolumn{3}{c|}{CC} & \multicolumn{3}{c|}{TF} & \multicolumn{3}{c|}{CC} & \multicolumn{3}{c|}{TF}  \\
\hline
\makecell{Classifier} & $H_{1}$ & $H_{2}$& $H_{1}$-$H_{2}$ & $H_{1}$ & $H_{2}$& $H_{1}$-$H_{2}$& $H_{1}$ & $H_{2}$& $H_{1}$-$H_{2}$& $H_{1}$ & $H_{2}$& $H_{1}$-$H_{2}$\\
\hline
SVM		&$\SI{80.8}{\percent}$   & $\SI{77.2}{\percent}$   & $\SI{83.1}{\percent}$   & $\SI{89.4}{\percent}$   & $\SI{83.2}{\percent}$   & \cellcolor[rgb]{0.13,0.67,0.8}$\SI{90.8}{\percent}$   & $\SI{81.0}{\percent}$   & $\SI{77.4}{\percent}$& $\SI{82.8}{\percent}$   & $\SI{89.2}{\percent}$   & $\SI{83.4}{\percent}$   & \cellcolor[rgb]{0.13,0.67,0.8}$\SI{90.9}{\percent}$  \\
LR	  &$\SI{76.0}{\percent}$   & $\SI{72.2}{\percent}$   & $\SI{75.4}{\percent}$   & $\SI{83.7}{\percent}$   & $\SI{77.5}{\percent}$   & $\SI{85.7}{\percent}$   & $\SI{76.2}{\percent}$   & $\SI{72.3}{\percent}$& $\SI{75.1}{\percent}$   & $\SI{83.8}{\percent}$   & $\SI{77.2}{\percent}$   & $\SI{85.6}{\percent}$  \\
RF    &$\SI{76.9}{\percent}$   & $\SI{75.0}{\percent}$   & $\SI{77.5}{\percent}$   & $\SI{88.9}{\percent}$   & $\SI{82.4}{\percent}$   & $\SI{89.6}{\percent}$   & $\SI{76.6}{\percent}$   & $\SI{75.1}{\percent}$& $\SI{77.1}{\percent}$   & $\SI{88.7}{\percent}$   & $\SI{82.2}{\percent}$   & $\SI{89.9}{\percent}$  \\
GB    &$\SI{88.3}{\percent}$   & $\SI{79.1}{\percent}$   & $\SI{89.5}{\percent}$   & $\SI{89.0}{\percent}$   & $\SI{82.7}{\percent}$   & $\SI{90.2}{\percent}$   & $\SI{88.5}{\percent}$   & $\SI{79.0}{\percent}$& $\SI{89.5}{\percent}$   & $\SI{88.7}{\percent}$   & $\SI{83.0}{\percent}$   & $\SI{90.5}{\percent}$  \\
\hline
\end{tabular}
}
\end{table*}
\begin{figure*}[hbt!]
\centering
\includegraphics[width=0.9\textwidth,height=0.7\textheight,keepaspectratio]{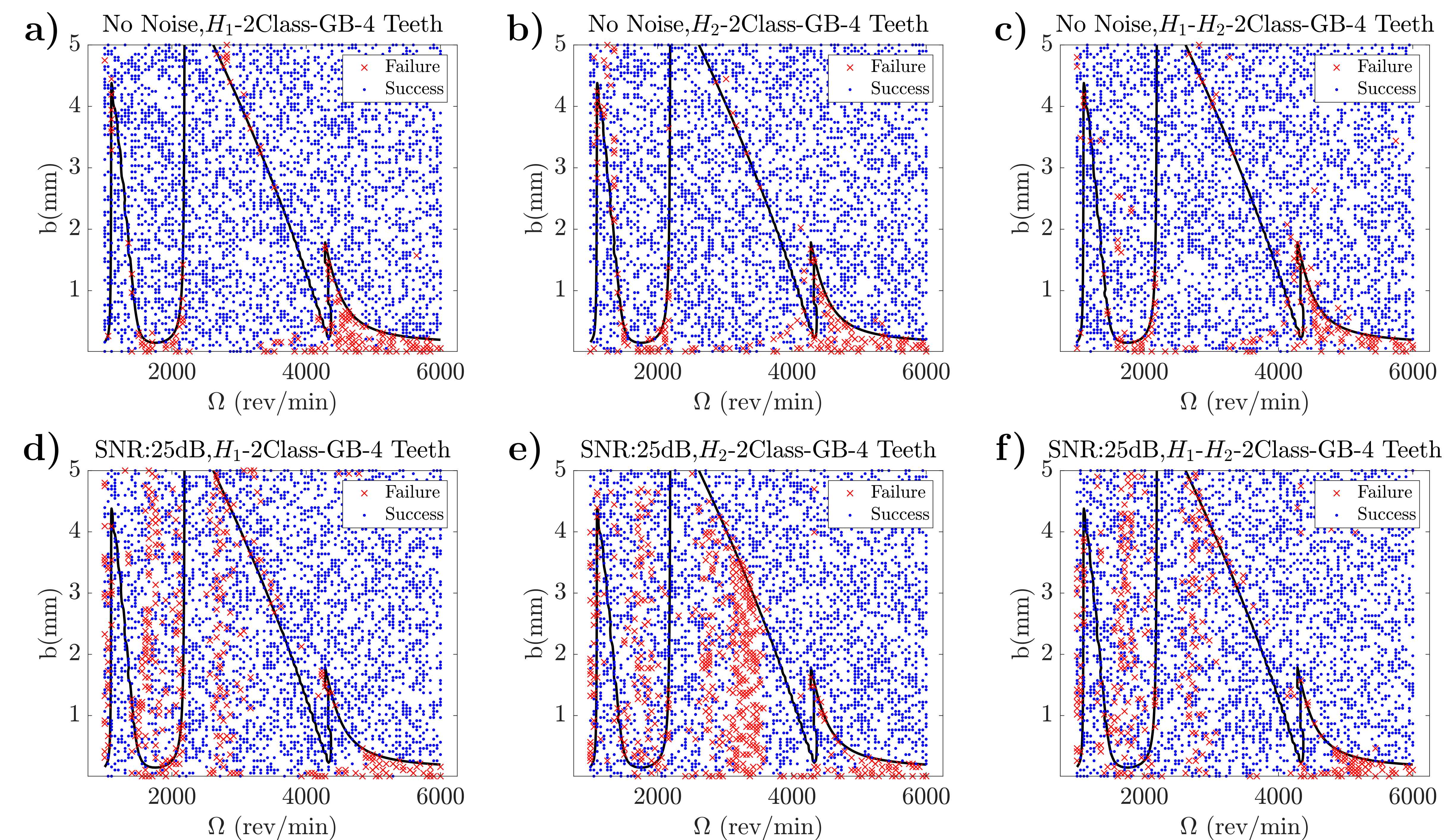}
\caption{Success and failure of two class classifications performed with Template Function feature matrices and Gradient Boosting algorithm for test set of data set without noise and with SNR value of 25 dB. a)Classification with 1D persistence features for non-noisy data set, b)Classification with 2D persistence features for non-noisy data set, c)Classification with 1D-2D persistence combined features for non-noisy data set, d)Classification with 1D persistence features for noisy data set with SNR:25 dB, e)Classification with 2D persistence features for noisy data set with SNR:25 dB, f)Classification with 1D-2D persistence combined features for noisy data set with SNR:25 dB.}
\label{fig:N4_downmill_SucFail_No_Noise_and_25dB}
\end{figure*}
\begin{table*}[t]
\centering
\caption{2 Class classification results for noisy (SNR:20,25,30 dB) and non-noisy data sets which belong to upmilling process with $N=4$ (N: Teeth Number, CC: Carlsson Coordinates, TF: Template Functions, SVM: Support Vector Machine, LR: Logistic Regression, $H_{1}$: 1D persistence, $H_{2}$: 2D persistence).}
\label{tab:Up_N4_CC_TF_2Class_Results}
\resizebox{\textwidth}{!} {  
\begin{tabular}{|l|c|c|c|c|c|c|c|c|c|c|c|c|}
\hline
\multicolumn{1}{|c|}{UpMilling} & \multicolumn{6}{c|}{\shortstack{Without Noise}} & \multicolumn{6}{c|}{\shortstack{SNR: 20 dB}} \\
\hline
\makecell{$N=4$} & \multicolumn{3}{c|}{CC} & \multicolumn{3}{c|}{TF} & \multicolumn{3}{c|}{CC} & \multicolumn{3}{c|}{TF}  \\
\hline
\makecell{Classifier} & $H_{1}$ & $H_{2}$& $H_{1}$-$H_{2}$ & $H_{1}$ & $H_{2}$& $H_{1}$-$H_{2}$& $H_{1}$ & $H_{2}$& $H_{1}$-$H_{2}$& $H_{1}$ & $H_{2}$& $H_{1}$-$H_{2}$\\
\hline
SVM		&$\SI{86.0}{\percent}$   & $\SI{78.5}{\percent}$   & $\SI{85.8}{\percent}$   & $\SI{86.1}{\percent}$   & $\SI{80.4}{\percent}$   & $\SI{86.0}{\percent}$   & $\SI{76.8}{\percent}$   & $\SI{80.7}{\percent}$& $\SI{82.1}{\percent}$   & $\SI{84.6}{\percent}$   & $\SI{84.0}{\percent}$   & \cellcolor[rgb]{0.13,0.67,0.8}$\SI{85.1}{\percent}$  \\
LR	  &$\SI{85.3}{\percent}$   & $\SI{77.8}{\percent}$   & $\SI{85.3}{\percent}$   & $\SI{86.2}{\percent}$   & $\SI{81.3}{\percent}$   & $\SI{85.8}{\percent}$   & $\SI{69.4}{\percent}$   & $\SI{80.4}{\percent}$& $\SI{80.9}{\percent}$   & $\SI{82.7}{\percent}$   & $\SI{81.3}{\percent}$   & $\SI{84.1}{\percent}$  \\
RF    &$\SI{84.9}{\percent}$   & $\SI{80.3}{\percent}$   & $\SI{84.9}{\percent}$   & $\SI{85.9}{\percent}$   & $\SI{81.2}{\percent}$   & $\SI{85.7}{\percent}$   & $\SI{75.5}{\percent}$   & $\SI{80.7}{\percent}$& $\SI{81.1}{\percent}$   & $\SI{82.8}{\percent}$   & $\SI{81.8}{\percent}$   & $\SI{83.2}{\percent}$  \\
GB    &\cellcolor[rgb]{0.13,0.67,0.8}$\SI{86.1}{\percent}$   & $\SI{80.9}{\percent}$   & $\SI{86.0}{\percent}$   & $\SI{85.6}{\percent}$   & $\SI{81.3}{\percent}$   & $\SI{86.0}{\percent}$   & $\SI{80.6}{\percent}$   & $\SI{82.2}{\percent}$& $\SI{82.5}{\percent}$   & $\SI{84.1}{\percent}$   & $\SI{83.4}{\percent}$   & $\SI{84.6}{\percent}$  \\
\hline
\multicolumn{1}{|c|}{Upmilling} & \multicolumn{6}{c|}{\shortstack{SNR: 25 dB}} & \multicolumn{6}{c|}{\shortstack{SNR: 30 dB}} \\
\hline
\makecell{$N=4$} & \multicolumn{3}{c|}{CC} & \multicolumn{3}{c|}{TF} & \multicolumn{3}{c|}{CC} & \multicolumn{3}{c|}{TF}  \\
\hline
\makecell{Classifier} & $H_{1}$ & $H_{2}$& $H_{1}$-$H_{2}$ & $H_{1}$ & $H_{2}$& $H_{1}$-$H_{2}$& $H_{1}$ & $H_{2}$& $H_{1}$-$H_{2}$& $H_{1}$ & $H_{2}$& $H_{1}$-$H_{2}$\\
\hline
SVM		&$\SI{85.3}{\percent}$   & $\SI{83.4}{\percent}$   & $\SI{84.8}{\percent}$   & \cellcolor[rgb]{0.13,0.67,0.8}$\SI{85.5}{\percent}$   & $\SI{84.4}{\percent}$   & \cellcolor[rgb]{0.13,0.67,0.8}$\SI{85.5}{\percent}$   & $\SI{83.2}{\percent}$   & $\SI{71.6}{\percent}$& $\SI{83.1}{\percent}$   & $\SI{85.9}{\percent}$   & $\SI{75.0}{\percent}$   & \cellcolor[rgb]{0.13,0.67,0.8}$\SI{86.2}{\percent}$  \\
LR	  &$\SI{79.2}{\percent}$   & $\SI{84.0}{\percent}$   & $\SI{84.1}{\percent}$   & $\SI{84.2}{\percent}$   & $\SI{84.5}{\percent}$   & $\SI{84.5}{\percent}$   & $\SI{79.4}{\percent}$   & $\SI{72.3}{\percent}$& $\SI{79.5}{\percent}$   & $\SI{84.1}{\percent}$   & $\SI{75.3}{\percent}$   & $\SI{85.0}{\percent}$  \\
RF    &$\SI{83.8}{\percent}$   & $\SI{84.1}{\percent}$   & $\SI{84.5}{\percent}$   & $\SI{83.5}{\percent}$   & $\SI{82.6}{\percent}$   & $\SI{83.0}{\percent}$   & $\SI{84.3}{\percent}$   & $\SI{75.0}{\percent}$& $\SI{84.4}{\percent}$   & $\SI{83.9}{\percent}$   & $\SI{75.3}{\percent}$   & $\SI{83.8}{\percent}$  \\
GB    &$\SI{85.2}{\percent}$   & $\SI{84.3}{\percent}$   & $\SI{84.8}{\percent}$   & $\SI{85.1}{\percent}$   & $\SI{84.5}{\percent}$   & $\SI{84.9}{\percent}$   & $\SI{85.1}{\percent}$   & $\SI{74.6}{\percent}$& $\SI{85.1}{\percent}$   & $\SI{85.7}{\percent}$   & $\SI{75.7}{\percent}$   & $\SI{85.2}{\percent}$  \\
\hline
\end{tabular}
}
\end{table*}


\begin{figure*}[hbt!]
\centering
\includegraphics[width=1\textwidth,height=0.7\textheight,keepaspectratio]{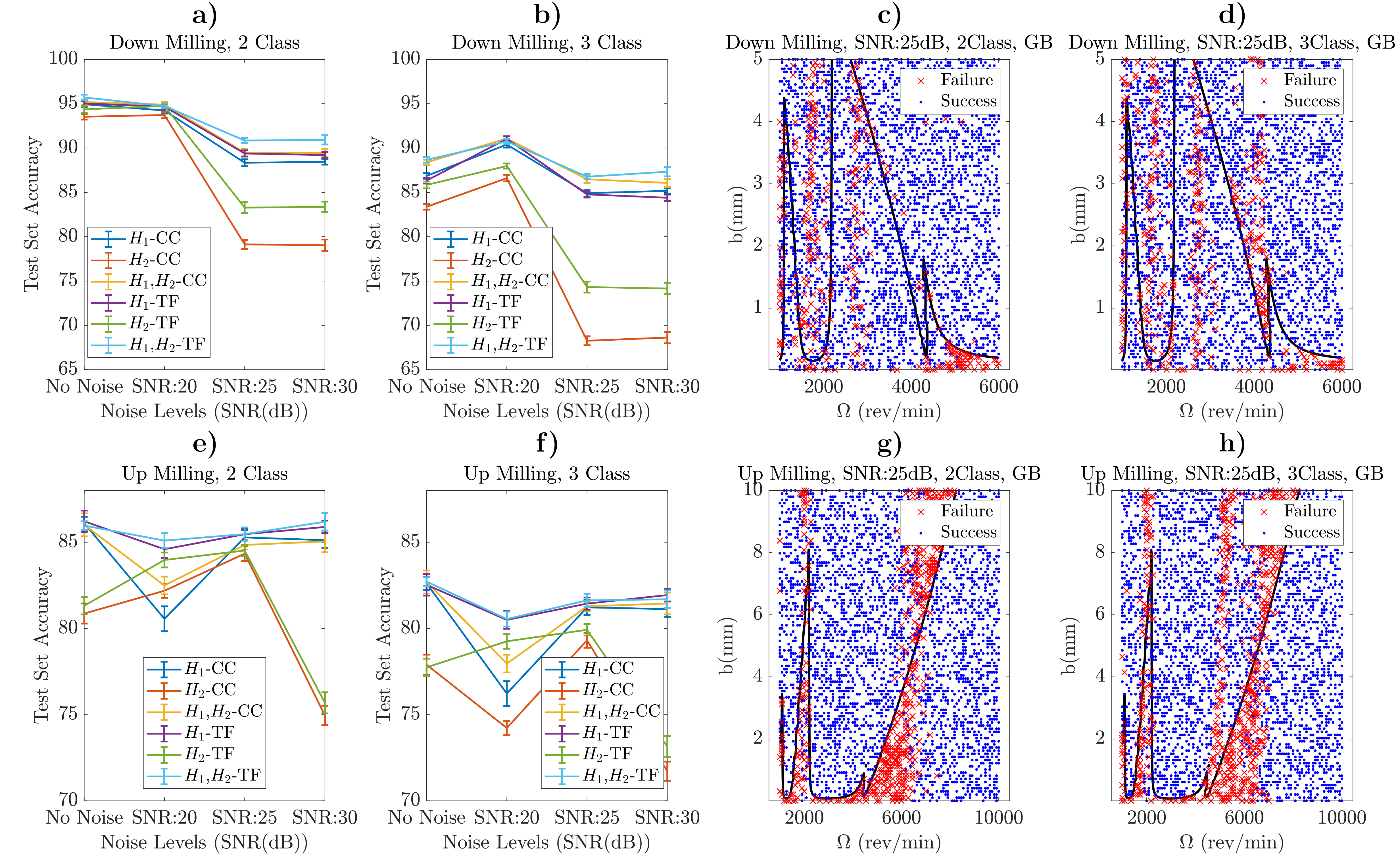}
\caption{Mean accuracies of downmilling process (a,b) and upmilling process (f,g) obtained for two class and three class classification performed with Carlsson Coordinates and Template Functions for non-noisy and noisy data sets where teeth number is 4. Two class(c) and three class (d) classification results obtained with Gradient Boosting algorithm is shown on the stability diagram for downmilling simulation data set whose SNR is 25 dB.Two class(e) and three class (h) classification results obtained with Gradient Boosting algorithm is shown on the stability diagram for upmilling simulation data set whose SNR is 25 dB.}
\label{fig:up_mill_down_mill_result_comp}
\end{figure*}

To compare the results of different levels of noise and different dimensions of persistence diagrams, classification results are plotted on the $100\times 100$ grid of stability diagram for milling process.
Figure~\ref{fig:N4_downmill_SucFail_No_Noise_and_25dB} presents the stability diagrams belongs to teeth number $N=4$ of down milling process for noisy data with SNR value of 25 dB and non-noisy data sets.
Figures on the first and second column belong to the classifications performed with only $H_{1}$ and $H_{2}$ features, respectively, while the ones in the third column represent the results of combinations of $H_{1}$ and $H_{2}$ features.
Red crosses on the stability diagrams denote the case that the prediction of the classifier does not match with the true label of the corresponding time series while blue dots shows matching between predictions and true labels.
From the figures, it is clear that the number of misclassifications increase slightly when the noise is introduced into the simulation data.
This is also reflected in Table~\ref{tab:Down_N4_CC_TF_2Class_Results} in the decrease in accuracies for different levels of noise, especially the noisy data sets with SNR value of 25 and 30 dB.

In addition, there is small accuracy difference which is at most $5\%$ between noisy (SNR:25, 30 dB) and non-noisy data set for downmilling cases, while this difference is less for upmilling results presented in Table~\ref{tab:Up_N4_CC_TF_2Class_Results}
This suggests that the featurization methods used yield promising results even with noisy data.
Persistent homology is known to be very robust against noise, as noise only adds points close to the diagonal which have short lifetimes.
Thus, these points do not contribute singificantly to the Carlsson coordinate or template function methods, making both featurizations robust against noise as well.

Figure~\ref{fig:up_mill_down_mill_result_comp} shows a comparison of the results obtained for up and downmilling with respect to different noise levels.
Since the deviations of accuracies for both featurization methods are relatively low, the classification accuracies can be considered reliable.
This trend is noticeable for both up and downmilling and for all levels of noise.
However, the classification results for upmilling are noticeably lower than those for downmilling.
Additionally, it is clear that $H_2$ features do not perform as well due to the lack of higher dimensional topological structure, as was explained earlier.
Figure~\ref{fig:up_mill_down_mill_result_comp} also presents the classification results on the stability diagrams for upmilling and downmilling for noisy data set.
It is seen that many misclassifications occur nearby the boundary of the stability diagram, especially for up milling process.
This boundary separates the unstable (above the boundary curve) and stable (under the boundary curve) cases so misclassifications on this boundary are likely.
It is also clear that when increasing to the three class problem, there are an increased number of misclassification.
However, the overall accuracy difference between both implementations, two and three class classification, does not exceed 5\% at their maximum accuracies.

The findings of this study indicate that topological features of data are appropriate descriptors for chatter recognition in milling. 
One advantage of the described approach is its ability to provide promising results without the need for manual preprocessing not only for non-noisy data sets, but also for time series with noise.
Therefore, we believe future work can include studying the effect of changing the simulation parameters on classification accuracy as well as experimental studies.

\section*{Acknowledgement}
This material is based upon work supported by the National Science Foundation under Grant Nos. CMMI1759823 and DMS-1759824 with PI FAK. The work of ST and EM was supported in part by NSF grants DMS-1800446, CMMI-1800466, and CCF-1907591.
\bibliography{ICMLA2019}
\newpage
\appendixpage
\label{sec:appendix}


\end{document}